\documentclass[review,5pt,twocolumn,preprint]{elsarticle}
\usepackage{lineno,hyperref}
\modulolinenumbers[5]
\usepackage{amsmath}

\begin{document}
\twocolumn[{\begin{frontmatter}
%\begin{frontmatter}

\title{Electronic Structure, Phase Stability and Resistivity of Hybrid Hexagonal C$_x$(BN)$_{1-x}$ Two-dimensional Nanomaterial:
A First-principles Study}

%\author{Ransell D'Souza, Sugata Mukherjee}
%\address{S.N. Bose National Centre for Basic Sciences, Block JD, Sector III, Salt Lake, Kolkata 700098, India}
\author[mymainaddress]{Ransell D'Souza}
\ead{ransell.d@gmail.com}

\author[mymainaddress]{Sugata Mukherjee\corref{mycorrespondingauthor}}
\cortext[mycorrespondingauthor]{Corresponding author}
\ead{sugata@bose.res.in; sugatamukh@gmail.com}

\address[mymainaddress]{S.N. Bose National Centre for Basic Sciences, Block JD, Sector III, Salt Lake, Kolkata 700098, India}

\begin{abstract}
We use density functional theory based first-principles method to  investigate the bandstructure and phase stability
in the laterally grown hexagonal C$_x$(BN)$_{1-x}$,  two-dimensional Graphene and {\it h}-BN hybrid nanomaterials, which were
synthesized by experimental groups recently (Liu {\it et al}, Nature Nanotech, {\bf 8}, 119 (2013)).  Our detail electronic structure
calculations on such materials, with both armchair and
zigzag interfaces between the Graphene and {\it h}-BN domains, indicate that the band-gap decreases non-monotonically with the concentration
of Carbon. The calculated bandstructure shows the onset of Dirac cone like features near the band-gap at high Carbon concentration
($x \sim 0.8$).  From the calculated energy of formation, the phase stability of C$_x$(BN)$_{1-x}$ was studied using a regular solution model
and the system was found to be in the ordered phase below a few thousand Kelvin. Furthermore, using the Boltzmann transport theory we calculate the
electrical resistivity from the bandstrcture of C$_x$(BN)$_{1-x}$ at different temperature ($T$), which shows a linear behaviour when plotted
in the logarithmic scale against $T^{-1}$, as observed experimentally..

\end{abstract}

%\begin{keyword}
%\texttt{elsarticle.cls}\sep \LaTeX\sep Elsevier \sep template
%\MSC[2010] 00-01\sep  99-00
%\end{keyword}

\end{frontmatter}
}]
\section*{Introduction}
Ever since the discovery of Graphene by Geim, Novoselov and co-workers \cite{geim04,geim07}, a great deal of effort has been put to make Graphene functionalized by engineering a controllable band-gap between its valence and conduction bands. Hexagonal Boron Nitride
({\it h}-BN), having lattice constant very close to that of Graphene and an insulating band-gap of nearly 5\-eV, which can be easily
synthesized  in the form of monolayer flakes \cite{hBN06, hBN07}, provides a wide range of possibilities to mix with Graphene to yield a varying
band-gap material depending on the degree of mixing \cite{geim09}. As such materials are of great importance in optoelectronic devices, a great deal of effort has been made to synthesize  hexagonal CBN ({\it h}-CBN) monolayer and multilayer nanomaterials
\cite{cnrrao09, ajayan10, dean10, levendorf12} with varying concentration of C / BN.

Experimentally {\it h}-CBN was synthesized initially by Panchakarla et al \cite{cnrrao09} and by Ci et al \cite{ajayan10} using Chemical Vapour
Deposition (CVD) technique,
where concentration of C or BN was carefully controlled. All {\it h}-CBN samples were reported to exhibit semiconducting behavior showing a
band-gap varying between a few meV to nearly an eV, a fact which has been verified by first-principles calculations \cite{ajayan13a}.
Formation of a band-gap in Graphene, when doped by Boron and Nitrogen, has been studied earlier \cite{kan,ding,dutta,pruneda,bhowmick,liu}. It was shown that upon
Boron (hole) doping the Dirac cone in Graphene is moved above the Fermi level and a gap appears, whereas upon Nitrogen (electron)
doping the Dirac cone is moved below the Fermi level \cite{sm12}. Upon co-doping of Graphene by both B and N a gap appears between the conduction
and valence bands making {\it h}-CBN a semiconductor where the band gap depends sensitively on the degree of doping and also on the thickness
of the layer.

Very recently, Liu et al \cite{ajayan13} have reported on synthesis of in-plane laterally grown heterostructures of Graphene and {\it h}-BN where
these two materials are seamlessly integrated lithographically with varying domain sizes. This
astounding synthesis of laterally grown hybrid C$_x$(BN)$_{1-x}$ two-dimensional heterostructure, with domain shapes such as circular
dots, stripes and patterns etc of varying sizes and width, has made the possibility of device application of such materials a reality.
Similar synthesis of hybrid {\it h}-CBN have been recently carried out by other groups \cite{gao,gang,liu14} using different experimental conditions.

Several calculations have been reported on the electronic structure
\cite{kan,ding,dutta,pruneda,bhowmick,liu,yu11,peng12} and also on the stability of C$_x$(BN)$_{1-x}$ \cite{jena11}. In this paper we calculate the
phase stability of CBN from the free energy using a regular solution model and apply the transport theory of band electrons on our DFT
bandstructure to obtain the temperature dependent resistivity at different concentration of C$_x$(BN)$_{1-x}$, which were not been addressed before.

Since, the interface between such domains can be either armchair  or zigzag type, we have studied both the interfaces using a large $5 \times 5$ unit cell. Recently, few such calculations \cite{grossman12,cnrrao13} have been reported, but these
authors have defined different unit cell types for armchair and zigzag interfaces.  In the present calculation the unit
cells for both armchair and zigzag interfaces are consistently similar giving rise to hexagonal Brillouin zone for  each of these interfaces.
We have consistently varied the concentration of C (or BN) as 0.2, 0.4, 0.6 and 0.8 and have calculated the band structure, density of states,
charge density and formation energy of C$_x$(BN)$_{1-x}$. Our calculated band structure for each of those interfaces show the emergence
of Dirac-cone like features with increasing C concentration.  {\it h}-BN has a band gap of nearly 5 eV whereas Graphene has zero band gap
at the high symmetric K-point in the hexagonal Brillouin zone, so it is expected for the band gap to decrease with the increasing C concentration,
ultimately becoming zero for $x=1$.
We obtained a non-monotonic decrease of the band gap for C$_x$(BN)$_{1-x}$
with increasing $x$ and the concentration dependence of the band gap is different for the armchair and zigzag interfaces. Our calculated
DOS and charge density indicate that the charge transfer effects might play important role in the band gap formation. Moreover, from the
calculated formation energy, we studied the phase stability of C$_x$(BN)$_{1-x}$ using a regular solution model and estimated the order-disorder
transition temperature. It was found that the onset of substitutional disorder would occur at temperatures of $\sim 3850$\-K and 6090\-K for
the zigzag and armchair interfaces, respectively.

Finally, we use the Boltzmann transport theory applied to the band electrons \cite{ashcroft76} to calculate the electrical conductivity ($\sigma$)\cite{madsen06}
from the bandstructure of C$_x$(BN)$_{1-x}$ calculated earlier. From the calculated $\sigma(T)$ we obtain the resistivity $\rho(T)$,
which shows a linear behaviour when plotted in the logarithmic scale against $T^{-1}$, as expected for semiconductors and as measured experimentally for C$_x$(BN)$_{1-x}$
\cite{ajayan10,ajayan13a}.

\section*{Method of Calculation}

We carried out the {\it ab-initio} DFT calculations on the $5 \times 5$ {\it h}-CBN unit-cell with armchair and zigzag (Fig. 1) interfaces using the Quantum Espresso code \cite{giannozzi09}.   A hexagonal unit cell was chosen for both the armchair and zigzag case.
The plane wave calculations  assume periodicity and hence to avoid interactions between the sheets, a vacuum spacing of 13\-$\mathring {\rm A}$ was chosen. We have used the ultrasoft pseudopotential
\cite{vanderbilt90} to describe the core electrons and the generalized gradient approximation (GGA) for the exchange-correlation kernel \cite{pbe96}. A kinetic energy cutoff of 40\-Ry was used for the plane-wave basis set and of 160\-Ry for the charge density, and an accuracy of $10^{-9}$\,Ry was obtained in the
self-consistent calculation of total energy. The equilibrium
lattice constants were obtained by minimizing the total energy with respect to the lattice constants by ensuring that the stress on
each of the atoms are zero. The self-consistent calculations were performed  using a converged Monkhorst-Pack $k$-point grids \cite{mp76} of
$6 \times 6 \times 1$. Band structure calculations were performed for the equilibrium lattice constants with 150 $k$-points along
the high-symmetric points $\Gamma$-K-M-$\Gamma$ in the
irreducible hexagonal Brillouin zone for both armchair and zigzag interfaces.

\begin{figure}[!ht]
\includegraphics[scale=0.3]{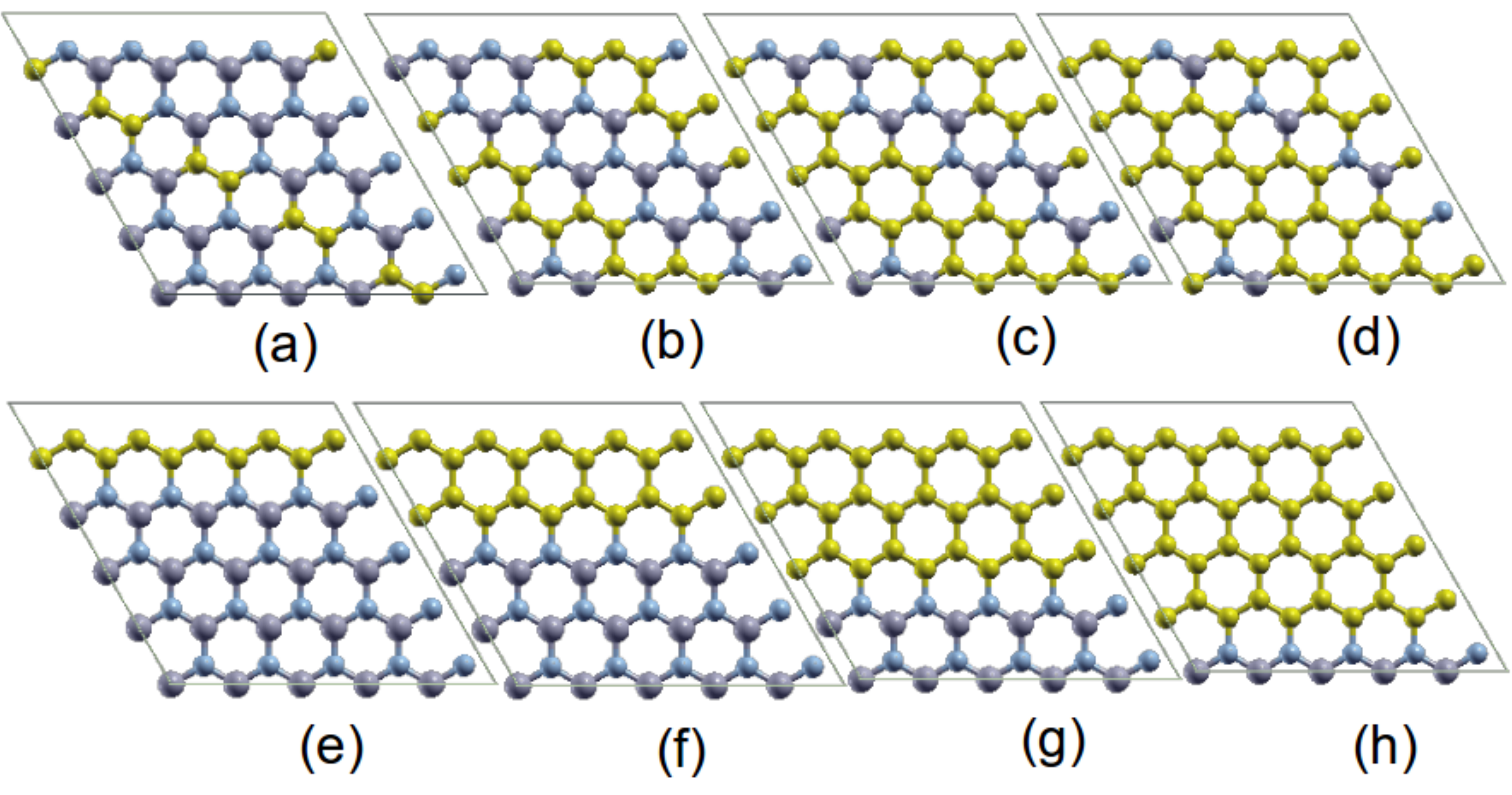}% Here is how to import EPS art
\caption{\label{fig:wide} $5 \times 5$ unit-cell of C$_x$(BN)$_{1-x}$ with armchair interface (a, b, c, d) and zigzag interface (e, f, g, h),
between Graphene and {\it h}-BN domains, at $x$ = 0.2, 0.4, 0.6 and 0.8; respectively. Carbon atoms are denoted by
yellow, Boron by grey and Nitrogen by blue colored balls; respectively.}
\end{figure}

We obtained the in-plane lattice constant
$a_0$ as 2.466\;$\mathring{\rm A}$ and 2.508\;$\mathring{\rm A}$ for Graphene and {\it h}-BN, respectively, which compare well with
the experimental values.
For C$_x$(BN)$_{1-x}$ system the equilibrium lattice constant lies in between that of Graphene and {\it h}-BN and it was calculated
for different concentrations for each interface types.  The total density of states (DOS) and the partial densities of states (PDOS)
projected on each orbital on the in-equivalent atoms in the unit cell were calculated for all concentrations and the L{\"o}wdin charge \cite{soler95} is
also obtained. The Xcrysden code \cite{xcrysden03} was used for visualizing the valence charge densities.

For the calculation of the resistivity $\rho(T)$ of C$_x$(BN)$_{1-x}$ from the calculated bandstructure we use the Boltzmann transport theory for band electrons
as implemented in the code BoltzTrap \cite{madsen06}. This calculation involves the evaluation of the electron group velocity $v_{\alpha}(i,{\bf k})$ referring
to $i$-th energy band and the $\alpha$-th component of the wave vector {\bf k}, from the band dispersion $\varepsilon(i,{\bf k})$, expressed as,
\begin{eqnarray}
v_{\alpha}(i,{\bf k}) &=& {1 \over \hbar} {\partial \varepsilon(i,{\bf k}) \over \partial k_{\alpha}}.
\end{eqnarray}
The electrical conductivity tensor is then obtained from,
\footnotesize{\begin{eqnarray}
{\sigma_{\alpha\beta}(T,\mu) \over \tau} &=& {1 \over V} \int e^2 v_{\alpha}(i,{\bf k})\, v_{\beta}(i,{\bf k}) [{-\partial f_\mu(T,\epsilon) \over \partial \epsilon}] d\epsilon,
\end{eqnarray}}
where $\mu$ is the chemical potential, $f_{\mu}$ is the Fermi-Dirac distribution function, $V$ is the volume of the sample, and $\tau$ is the Drude relaxation time
which is assumed to be isotropic and direction independent \cite{allen88,allen92}. The resistivity is then obtained from the conducivity as, $\rho = 1/\sigma$.
The relaxation time $\tau$
is typically $\sim 10^{-14}$s \cite{ashcroft76} but a precise value of this quantity is unknown for C$_x$(BN)$_{1-x}$. For the calculation of the resistivity
of C$_x$(BN)$_{1-x}$ we have taken a very dense k-point grid of $40\times40\times1$ and even denser grid was taken at some concentrations for the band structure
calculation using Quantum Espresso code, which was then fed into the BoltzTrap code for the transport calculation.

\section*{Results and Discussion}
\subsection*{(a) Electronic structure}

In Fig. 2 we show the bandstructure and the corresponding density of states (DOS) of C$_x$(BN)$_{1-x}$ for $x = 0.2, 0.4, 0.6$ and $0.8$, for both armchair and zigzag interfaces between the Graphene and {\it h}-BN domains. The DOS shown in Fig. 2 refers to total DOS (including $2s$, $2p_x$, $2p_y$ and $2p_z$ contributions) of Carbon, Boron and Nitrogen atoms in the unit-cell and also the partial density of states (PDOS) of $2p_z$ orbitals on each of those atoms. The DOS and PDOS were calculated for all nonequivalent atoms in the
unit-cell, so that in the figures they appear appropriately weighted. The Fermi energy is at the middle of the energy gap between the conduction and the valence bands. It should be recalled that $\pi$ and ${\pi}^*$ bands of carbon dominate the electronic structure of undoped Graphene around $E_F$, and ${\pi}^*$ bands of B and $\pi$ bands of N represent the conduction and valence bands of {\it h}-BN, above and below its energy gap \cite{sm11}, respectively. It is clear from the DOS results shown in Fig. 2 that the nature of the bands around the energy gap of C$_x$(BN)$_{1-x}$ are essentially due to $2p_z$ states of C, B and N, as the total DOS is completely dominated by PDOS of $2p_z$ state, around 2.5\-eV each above and below the band gap for all concentrations.

\begin{figure}[!ht]
\includegraphics[scale=0.3]{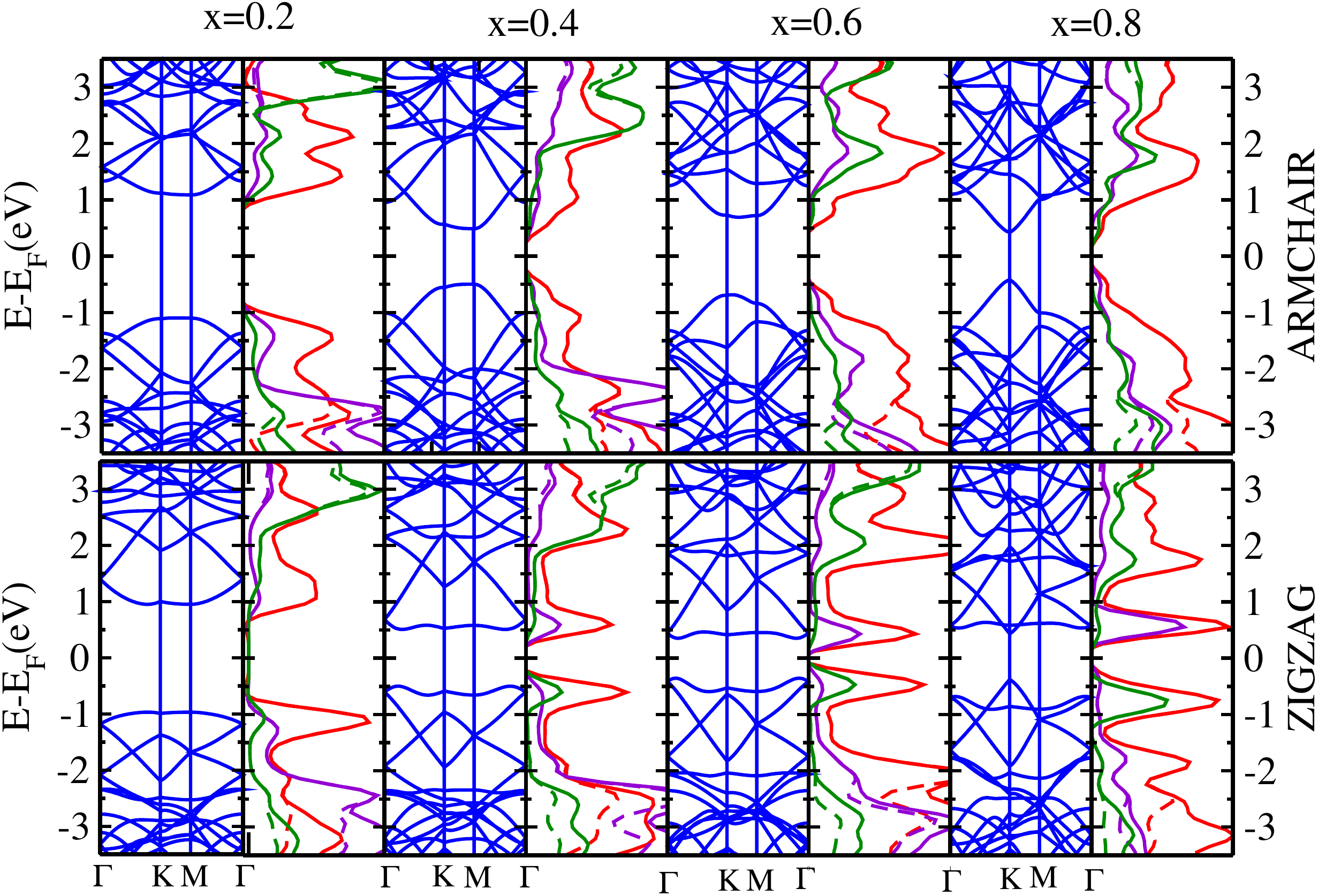}% Here is how to import EPS art
\caption{\label{fig:wide}  The calculated bandstructure and the density of states (DOS) of C$_x$(BN)$_{1-x}$ at $x=0.2, 0.4, 0.6$ and
$0.8$ for armchair
(upper panel) and zigzag (lower panel) interfaces. The Fermi energy is at the centre of the band gap. The bands (blue) are shown in
the high-symmetry directions $\Gamma$-K-M-$\Gamma$ in the hexagonal Brillouin zone. The total DOS is shown as full-line
for C (red), N (violet) and B (green); the corresponding projected density of states (PDOS) of $2p_z$ states for each atoms being shown as dashed-line with similar colors, respectively. The DOS and PDOS are in arbitrary units.}
\end{figure}
The bandstructure in Fig. 2 shows that the band gap of C$_x$(BN)$_{1-x}$ decreases with increasing concentration of C. Since the band gap of undoped {\it h}-BN is calculated to be 4.76\-eV, the band gap of C$_x$(BN)$_{1-x}$ decreases from this value with increasing C-concentration upon mixing with  Graphene until it becomes zero for $x=1$. The nature of the decrease of the band gap with increasing
$x$ was found to be non-monotonic and different for armchair and zigzag interfaces between Graphene and {\it h}-BN domains as shown in Fig 3., to be discussed later. For the armchair interface the minimum band gap appears near the high symmetric M-point of the hexagonal Brillouin zone for $x = 0.2$ and $0.4$. For $x = 0.6$, C$_x$(BN)$_{1-x}$ behaves as an indirect band gap material for the armchair interface, with the minima of the conduction band lying between the high symmetric K- and M-points. This behavior changes at higher C-concentration when at $x = 0.8$, Dirac cone-like feature appears at the K-point in both armchair and zigzag interfaces, as expected for Graphene. For the zigzag interface, C$_x$(BN)$_{1-x}$ behaves as
indirect gap material for $x = 0.2, 0.4$ and 0.6. At $x=0.8$ the direct band gap at K-point is very close to that of armchair case.
We would like to emphasize that all our self-consistent calculations were performed by relaxing the lattice, but keeping the hexagonal symmetry,
to assure zero strain in the unit cell. In this way, we have obtained the equilibrium in-plane lattice constant $a(x)$ for each concentration.
We would like to add that our calculations were performed for the spin polarized case. The spin up and spin down components of the DOS happen to be exactly the same unlike reported by \cite{kan,ding,dutta,pruneda,bhowmick,liu}. The reason for this is that those calculations are performed on nanoribbons. The spin polarization on each atom diminishes as we move into the nanoribbon \cite{charge}. In this paper all our calculations are studied on an infinite sheet and hence there are no spin polarization on any atom.
\begin{figure}[!ht]
\includegraphics[scale=0.3]{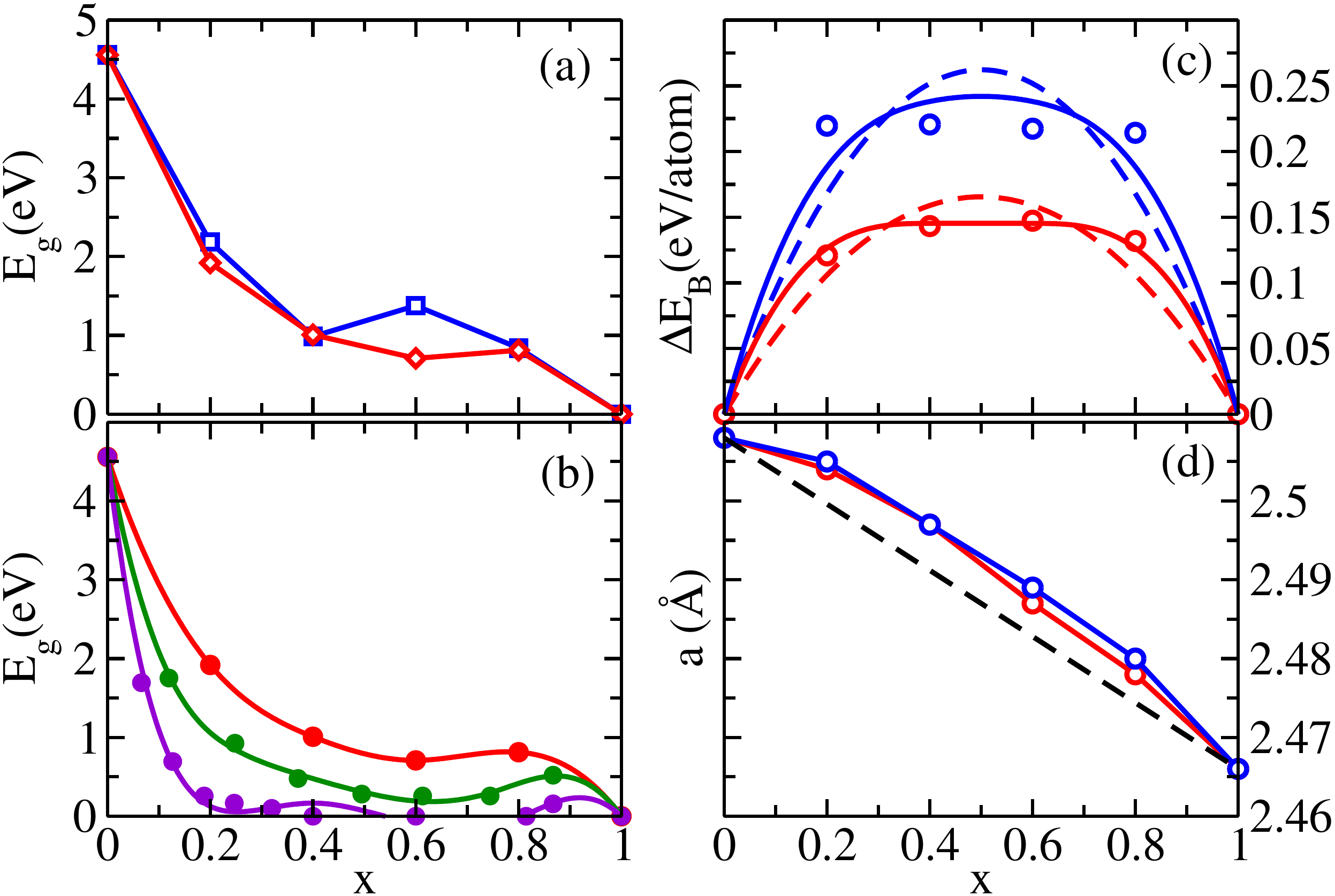}% Here is how to import EPS art
\caption{\label{fig:wide}  Various calculated physical properties of C$_x$(BN)$_{1-x}$ shown as a function of $x$, both for
armchair (blue) and zigzag (red) interfaces between Graphene and {\it h}-BN domains.
(a) The minimum band gap. (b) Band gap for the zigzag interface fitted with Eq (1), shown for present calculation (red), that of $8\times8$
cell (green) and that of $16\times16$ cell (violet) of Bernardi {\it et al} \cite{grossman12}.
(c) Formation energy $\Delta E_B$ for both type of interfaces. Calculated data are shown as open circles. Parabolic fit by functions
$4\,\Delta H\, x(1-x)$ are shown as dashed lines. A fit by a function of the form $H_0 + H_1\,x(1-x) + H_2\,x^2(1-x)^2$ are shown by
solid lines.
(d) The equilibrium in-plane lattice constant $a$. The dashed line refers to Vegard's law.}
\end{figure}

In Fig. 3 the calculated band gap $E_g$, the formation energy $\Delta E_B$, and the equilibrium lattice constant $a$ are shown for
C$_x$(BN)$_{1-x}$ as a function of $x$, for the armchair and zigzag interfaces.
The calculated $E_g$ indicates Fig. 3(a), for both armchair and zigzag interfaces, the indirect band gap to be slightly smaller than the
direct band gap at high-symmetric K-point for $x < 0.8$. For $x \geq 0.8$, the Dirac cone-like features start evolving and the
band gap is a direct one at K-point. This non-monotonic behavior of $E_g$, which has also been observed in recently published calculations
\cite{grossman12,cnrrao13}, is apparently not observed in the semiconductor alloys. This dependence can be incorporated by taking higher order
concentration dependent terms in the optical bowing parameter \cite{zunger87}, so we have fitted a fifth-order polynomial to
best describe the $x$ dependence of $E_g$ for C$_x$(BN)$_{1-x}$, given as,
\begin{eqnarray}
E_g(x) &=& [E_g^{\rm{\it h}BN} + E_0\, x + E_1\, x^2]\, (1-x) \nonumber \\
& & + [ E_2\, + E_3\, x]\, x^2(1-x)^2.
\end{eqnarray}
Here, $E_g^{\rm{\it h}BN}$ is the energy gap of undoped {\it h}-BN, $E_0$ is the optical bowing parameter, $E_1, E_2, E_3$ are the higher order corrections to the bowing 
parameter, obtained by the fitting procedure given in Table 1.
We observe that the concentration dependence of the band gap results of  Bernardi {\it et al} \cite{grossman12} performed for larger zigzag interface unit-cell, fit nicely with the form given in Eq (1) shown in Fig 3(b).

\subsection*{(b) Formation energy and phase stability}

The formation energy $\Delta  E_B$ Fig. 3(c) was calculated using the equation,
\begin{eqnarray}
\Delta  E_B &=& E\{{\rm C}_x({\rm BN})_{1-x}, a(x) \} \nonumber \\
& & -  [ x\- E({\rm C}, a_{\rm C}) + \ (1-x) \- E(h{\rm BN}, a_{h{\rm BN}})],  
\end{eqnarray}
where $E\{{\rm C}_x({\rm BN})_{1-x}, a(x) \}$ is the total energy per atom of C$_x$(BN)$_{1-x}$ at the equilibrium in-plane lattice constant
$a(x)$; $E({\rm C}, a_{\rm C})$ and $E({\rm{\it h}BN}, a_{h{\rm BN}})$ are the total
energies per atom of undoped Graphene and {\it h}-BN at the equilibrium in-plane lattice constants $a_{\rm C}$ and $a_{h{\rm BN}}$, respectively.

From the calculated formation energy, we investigated the phase stability \cite{lambrecht93, neugebauer95} of C$_x$(BN)$_{1-x}$ by
fitting the calculated data with a parabola, expressing $\Delta  E_B(x) = 4\,\Delta H\, x(1-x)$,
where $\Delta H$ is the formation energy at $x=0.5$. In the regular solution model, the entropy of mixing can be expressed as point probabilities or
the concentration $x$ as, $S = -k_B[x\,\ln x + (1-x)\, \ln(1-x)]$, where $k_B$ is the Boltzmann constant. The Free energy is expressed as $F(T,x) = \Delta E_B(x) - T\,S$. At low
temperatures $F(T,x)$ shows a maximum at $x=0.5$, and two minima located symmetrically away from  $x=0.5$. With increasing temperature these
two minima converge to give rise to a single minimum at a critical temperature $T_{C}$ at $x=0.5$. The critical temperature was obtained
from the equation,  fulfilling the condition that $d^2F/dx^2 < 0$ is unstable and bounded by the spinodal line, given by \cite{lambrecht93},
\begin{eqnarray}
k_B\-T &=& 8\,\Delta H \, x(1-x).
\end{eqnarray}
Thus, the critical temperature, $T_{C}=2\,\Delta H/k_B$, was estimated to be 3850\-K for the zigzag and 6090\-K for the armchair interfaces, respectively. Therefore it is expected C$_x$(BN)$_{1-x}$ would be in the disordered phase above those temperatures. A lower bound of
$T_{C}$ can be obtained by estimating $\Delta H$ directly from interpolation of the calculated $\Delta E_B$ at $x=0.5$, yielding the transition temperatures to be
3390\-K and 5060\-K for the zigzag and armchair interfaces, respectively.

We have also investigated the phase stability of C$_x$(BN)$_{1-x}$ by using the fit $\Delta E_B(x) = H_0 + H_1\,x(1-x) + H_2\,x^2(1-x)^2$, which gives better than parabolic
fit shown in Fig 3(c) as full lines. Inclusion of such higher order terms in $\Delta E_B(x)$ leads the transition from binodal to spinodal line to occur at
temperatures lower than in the previous model. The free energy was calculated numerically and the $T_{C}$ was found to be 4869\-K for the armchair and 3389\-K for the
zigzag interface \cite{charge}, respectively. We would like to mention that above calculations for larger supercells are under investigation and will be reported later.

%% Table begins from here
%\section*{Table}
\begin{table}[!htbp]%The best place to locate the table environment is directly after its first reference in text
\caption{\label{tab:table1}%
Numerical value of the parameters in Eqs (3), (5) and (6). $E_0, E_1, E_2, E_3$ are in eV, $\Delta H$ in eV/atom and $A$ in
$\mathring {\rm A}$, respectively. The values of $E_n$ without parenthesis refer to that of the band gap at K-point and those
within parenthesis to the indirect band gap, respectively.
}
\resizebox{0.48\textwidth}{!}{%\begin{tabular}{L{3em}C{3em}C{3em}C{3em}C{3em}C{3em}C{3em}}
\begin{tabular}{lcccccc}
\hline
\textrm{Interface}&
\multicolumn{1}{c}{\textrm{$E_0$}}&
\textrm{$E_1$} & \textrm{$E_2$} & \textrm{$E_3$} & \textrm{$\Delta H$} &\textrm{$A$} \\
\hline
Armchair & -5.0     & -3.993    & -48.698   & 116.768   &  0.262 &\ \  0.028 \\
         & (0.1813) & (-10.823) & (-76.328) & (158.151) &         \\
Zigzag   & -16.554  & 18.333    & 11.053    & -6.605    &  0.166 &\ \  0.022 \\
         & (-17.952)& (25.333)  & (23.568)  & (-52.396) &         \\
\hline
\end{tabular}}
\end{table}
%% table ends over here

The in-plane lattice constant of C$_x$(BN)$_{1-x}$, $a(x)$ shows a deviation from Vegard's law \cite{liou05} in Fig. 3(d), which has been
fitted to,
\begin{eqnarray}
a(x) &=& x\, a_{\rm C} + (1-x)\, a_{h{\rm BN}} + A\, x(1-x).
\end{eqnarray}
Here, $A$ is the deviation parameter for the lattice constant $a$, obtained from fitting. The fitting parameters in Eqs. (1), (3) and (4)
are given in Table 1.

\begin{figure}[!rho]
\includegraphics[scale=0.3]{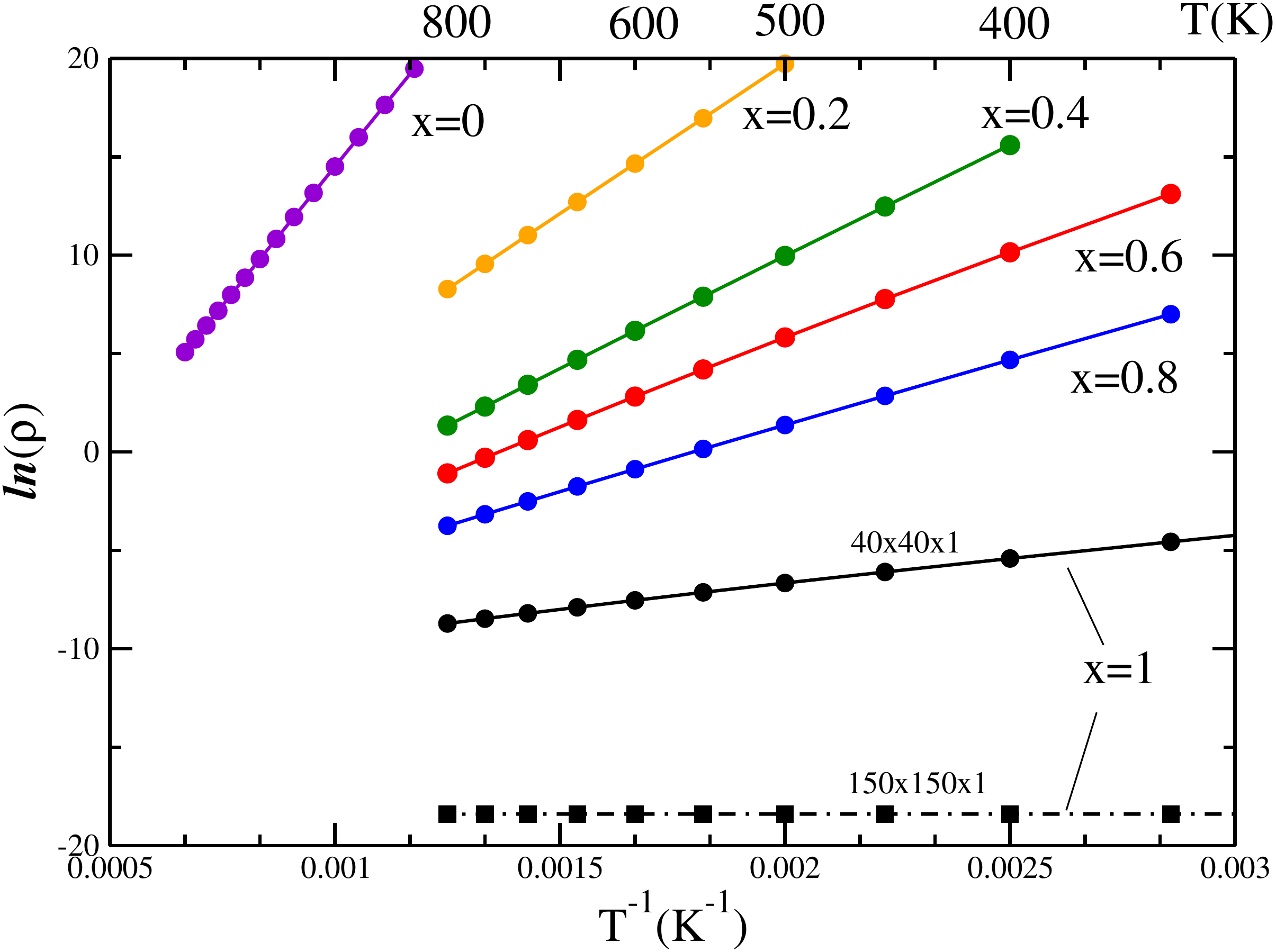}% Here is how to import EPS art
\caption{\label{fig:wide} $\ln(\rho(T))$ plotted against $T^{-1}$ for C$_x$(BN)$_{1-x}$ at different concentrations calculated from
the Boltzmann transport theory \cite{madsen06} at different concentration.}
\end{figure}

\subsection*{(c) Resistivity from the Transport theory}

Now we turn to our results of the resistivity $\rho(T)$ of C$_x$(BN)$_{1-x}$ from the transport calculations. The resistivity of CBN nanomaterials
was measured earlier \cite{ajayan10,ajayan13}. It was reported that $\ln(\rho)$ varies linearly with $T^{-1}$ for different concentration of B and N,
indicating that CBN is semiconducting. The band gap $E_g$ of CBN was estimated from Eq. (7) \cite{ashcroft76}. In Fig. (4) we show the results of $\ln(\rho)$
against $T^{-1}$, assuming $\tau = 10^{-14}$s, for the zigzag interface of C$_x$(BN)$_{1-x}$ at $x=0,0.2,0.4,0.6,0.8,1$, in the temperature range of 200K to 800K.
The calculated data can be fitted very well to straight lines as shown in Fig 4 and  the band gap $E_g$ at each concentration was calculated from the slope of
the lines using the relation,
\begin{eqnarray}
\rho(T) &=& \rho_{\infty} \exp(E_g/2k_BT). 
\end{eqnarray}
As mentioned earlier the
k-point mesh had to be enhanced to $150\times150\times1$ for pure Graphene ($x=1$) to capture the Dirac-point correctly. Also, for the pure {\it h}-BN we had
to calculate $\rho(T)$ at higher temperatures to obtain the measurable slope as shown in Fig. 4. In Table 2, the band gap $E_g$ of C$_x$(BN)$_{1-x}$ calculated
from the transport theory and those calculated directly using DFT are compared. We find an overall good agreement. It should be mentioned that the
numerical value of the relaxation time $\tau$ does not affect the the band gap estimation from the slope of Fig 4, since the constant $\rho_{\infty}$ in Eq (7) will
only shift the origin of the lines in Fig 4 and not affect their slopes. Any discrepancy of the calculated $E_g$ should thus come from inadequate k-point mesh.
To our knowledge, this is apparently the first calculation of $\rho(T)$ for the semiconducting nanomaterial C$_x$(BN)$_{1-x}$ from Boltzmann transport theory.

\begin{table}[b]%The best place to locate the table environment is directly after its first reference in text
\centering
\caption{\label{BSBT}%
Band gap calculated by Bandstructure and Boltzmann transport theory for C$_x$(BN)$_{1-x}$ for zigzag interface  at different concentrations.}
\footnotesize{\begin{tabular}{lcc}
\hline
$x$ & Bandstructure & Using Eq. (7) and Fig. (4) \\
\  & (Quantum Espresso) & (Boltztrap) \\
\hline
0 & 4.557 & 4.87 \\
0.2 & 1.919 & 2.63 \\
0.4 & 1.008 & 1.97 \\
0.6 & 0.709 & 1.53 \\
0.8 & 0.812 & 1.153 \\
\hline
\end{tabular}}
\end{table}

\subsection*{(d) Charge density and the PDOS}

Finally, in Fig. 5 we show the PDOS and the valence charge density on all in-equivalent atoms across the armchair (Figs. 4a and 4b)
and the zigzag (Figs 4c and 4d) interfaces of C$_x$(BN)$_{1-x}$ at $x=0.6$. The calculated PDOS give a idea about the contributions
coming from each in-equivalent C, B and N atoms towards the total DOS shown in Fig. 2. The calculated valence charge density (Figs 4b and 4d)
indicates that covalent $sp$-bonding nature is preserved in C$_x$(BN)$_{1-x}$.

\begin{figure}[!ht]
\includegraphics[scale=0.25]{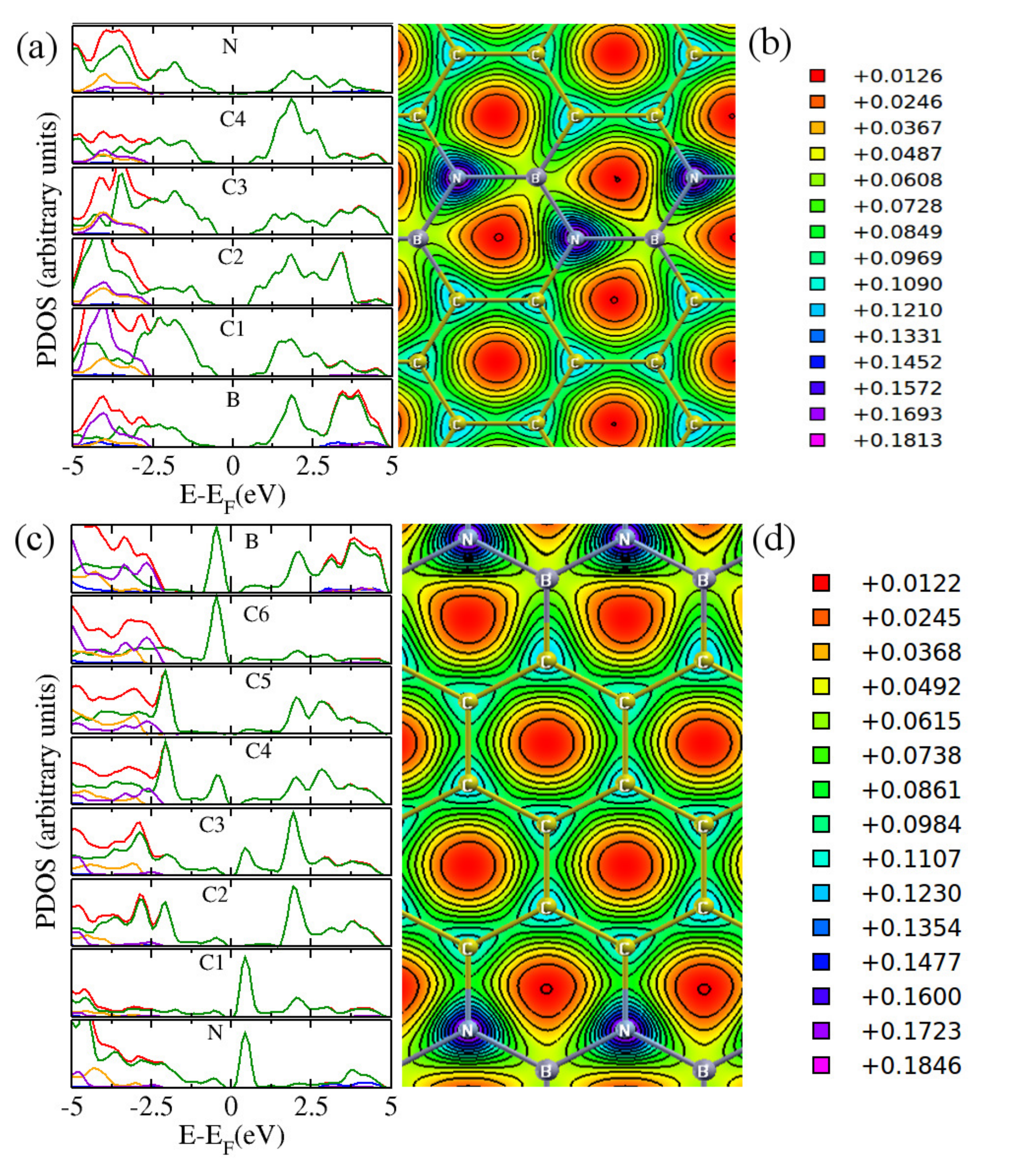}% Here is how to import EPS art
\caption{\label{fig:wide}  (a) : Calculated PDOS on the in-equivalent B, C and N atoms in the unit-cell. C1, C2, C3, C4 denote
four C-atoms
on the upper hexagon, terminated by B and N shown in (b). The PDOS referring to $2s$, $2p_{\rm total}, 2p_z, 2p_x$ and $2p_y$ orbitals are shown in blue, red, green, orange and violet, respectively.
(b) : Calculated valence charge density shown across the armchair interface between Graphene and {\it h}-BN domains.  The contours are in the units of $e/{\rm Bohr}^3$.
(c) : Calculated PDOS on the in-equivalent B, C and N atoms in the unit-cell. C1 ... C6 denote six C-atoms
on the chain, terminated by N and B atoms on two opposite zigzag interfaces shown in (d). (d) : Calculated valence charge density shown across the zigzag interface between Graphene and {\it h}-BN domains. (a) and (b) refer to C$_{0.6}$(BN)$_{0.4}$ armchair interface, whereas (c) and (d) refer to the zigzag interface. }
\end{figure}

The band structure and the DOS of C$_x$(BN)$_{1-x}$ are somewhat different for the zigzag interface than armchair interface.
The bands immediately above and below the energy gap are more flat, as evidenced by a strong peaks in the DOS. It should be
noted that unlike in the armchair interface, in zigzag interface the C atoms are terminated by either all B-atoms or by
all N-atoms (Fig. 1). This leads to different type of excess charge at the interfacial C-atoms. We have calculated this excess charge
from the difference of the L{\"o}wdin charges between the similar atoms in  C$_x$(BN)$_{1-x}$ and that of undoped Graphene and {\it h}-BN \cite{charge}.

We found, a C-atom terminated
by a B (N) atom at the zigzag interface would have more negative (positive) charge than that in undoped Graphene; whereas on the
zigzag interface on the other side of the same domain the excess charge on the interfacial C atom would be reversed. This
leads to strong peaks in the DOS above or below $E_F$, which alternates as one goes onto atoms lying deeper in the domain.
This effect is illustrated in Fig. 4c where we show the calculated PDOS on C-atoms going from one end of the zigzag interface to the
other end. Comparing the excess charges on the interfacial C atoms, for both armchair and zigzag interfaces, we found higher is this
excess charge, larger is the band gap $E_g$.

Present calculations may be extended to include higher order corrections
to the exchange-correlation energy using HSE \cite{hse} or GW \cite{gw} methods to check the validity of our results. However, the GGA exchange-correlation
kernel used in present calculations yields the groundstate physical properties of Graphene and {\it h}-BN, in good agreement with experimental results.

In conclusion, we have presented a detail first-principles calculation of the band structure, DOS, the band gap and the formation energy
of C$_x$(BN)$_{1-x}$
at different concentrations. From the formation energy, we have also investigated the phase stability of the material using a regular solution model.
Although we have used only the single-site probabilities for the entropy, which can be improved further by incorporating the pair or cluster probabilities,
we have given an estimate of the transition temperature for the order-disorder transition in C$_x$(BN)$_{1-x}$, apparently for the first time.
We have calculated the resistivity of C$_x$(BN)$_{1-x}$ using Boltzmann transport theory and have estimated the band gap of this semiconducting nanomaterial at
different concentrations which agrees with earlier experimental observations.
Our calculated DOS and PDOS should motivate further angle resolved photoemission spectroscopic (ARPES) measurements on this
technologically important nanomaterial.

\section*{Acknowledgements}
The calculations were performed at the High Performance Cluster parallel computer of S.N. Bose National Centre. RD would like to thank Prof T. Saha-Dasgupta for useful discussions.


\begin{thebibliography}{10}
\expandafter\ifx\csname url\endcsname\relax
  \def\url#1{\texttt{#1}}\fi
\expandafter\ifx\csname urlprefix\endcsname\relax\def\urlprefix{URL }\fi
\expandafter\ifx\csname href\endcsname\relax
  \def\href#1#2{#2} \def\path#1{#1}\fi

\bibitem{geim04}
K.~S. Novoselov, A.~K. Geim, S.~V. Morozov, D.~Jiang, Y.~Zhang, S.~V. Dubonos,
  I.~V. Grigorieva, A.~A. Firsov, Science 306 (2004) 666.

\bibitem{geim07}
A.~K. Geim, K.~S. Novoselov, Nature Mater 6 (2007) 183.

\bibitem{hBN06}
M.~Morscher, M.~Corso, T.~Greber, J.~Osterwalder, Surf. Sci. 600 (2006) 3280.

\bibitem{hBN07}
A.~Goriachko, Y.~He, M.~Knapp, H.~Over, M.~Corso, T.~Brugger, S.~Berner,
  J.~Osterwalder, T.~Greber, Langmuir 23 (2007) 2928.

\bibitem{geim09}
A.~Geim, Science 324 (2009) 1530.

\bibitem{cnrrao09}
L.~S. Panchakarla, K.~S. Subrahmanyam, S.~K. Saha, A.~Govindaraj, H.~R.
  Krishnamurthy, U.~V. Waghmare, C.~N.~R. Rao, Adv. Mater 21 (2009) 4726.

\bibitem{ajayan10}
L.~Ci, L.~Song, C.~Jin, D.~Jariwala, D.~Wu, Y.~Li, A.~Srivastava, Z.~F. Wang,
  K.~Storr, L.~Balicas, F.~Liu, P.~M. Ajayan, Nature Mater 9 (2010) 430.

\bibitem{dean10}
C.~R. Dean, A.F.Young, I.Meric, C.Lee, L.Wang, S.Sorgenfrei, K.Watanabe,
  T.Taniguchi, P.~Kim, K.L.Shepard, J.~Hone, Nature Nanotech. 5 (2010) 722.

\bibitem{levendorf12}
M.~Levendorf, C.~Kim, L.~Brown, P.~Huang, R.~Havener, D.~Muller, J.~Park,
  Nature 488 (2012) 627.

\bibitem{ajayan13a}
B.~Muchharla, A.~Pathak, Z.~Liu, L.~Song, T.~Jayasekera, S.~Kar, R.~Vajtai,
  L.~Balicas, P.~Ajayan, S.~Talapatra, N.~Ali, Nano Lett. 13 (2013) 3476.

\bibitem{kan}
E.~Kan, X.~Wu, Z.~Li, X.~Zeng, J.~Yang, J.~Hou, J Chem Phys 129 (2008) 084712.

\bibitem{ding}
Y.~Ding, Y.~Wang, J.~Ni, Applied Physics Letters 95 (2009) 123105.

\bibitem{dutta}
S.~Dutta, A.~K. Manna, S.~K. Pati, Physical Review Letters 102 (2009) 096601.

\bibitem{pruneda}
J.~M. Pruneda, Physical Review B 81 (2010) 161409(R).

\bibitem{bhowmick}
S.~Bhowmick, A.~K. Singh, B.~I. Yakobson, The Journal of Physical Chemistry C
  115 (2011) 9889.

\bibitem{liu}
Y.~Liu, S.~Bhowmick, B.~I. Yakobson, Nano Letter 11 (2011) 3113.

\bibitem{sm12}
S.~Mukherjee, T.~P. Kaloni, J Nanoparticle Res. 14 (2012) 1059.

\bibitem{ajayan13}
Z.~Liu, L.~Ma, G.~Shi, W.~Zhou, Y.~Gong, S.~Lei, X.~Yang, J.~Zhang, J.~Yu,
  K.P.Hackenberg, A.~Babakhani, J.~Idrobo, R.~Vajtai, J.~Lou, P.~Ajayan, Nature
  Nanotech 8 (2013) 119.

\bibitem{gao}
Y.~Gao, Y.~Zhang, P.~Chen, Y.~Li, M.~Liu, T.~Gao, D.~Ma, Y.~Chen, Z.~Cheng,
  X.~Qiu, W.~Duan, Z.~Liu, Nano Letters 13 (2013) 3439.

\bibitem{gang}
G.~H. Han, J.~A. Rodrıguez-Manzo, C.-W. Lee, N.~J. Kybert, M.~B. Lerner, Z.~J.
  Qi, E.~N. Dattoli, A.~M. Rappe, M.~Drndic, A.~T.~C. Johnson, ACS Nano 7
  (2013) 10129.

\bibitem{liu14}
L.~Liu, J.~Park, D.~A. Siegel, K.~F. McCarty, K.~W. Clark, W.~Deng, L.~Basile,
  J.~C. Idrobo, A.-P. Li, G.~Gu, Science 343 (2014) 163.

\bibitem{yu11}
Z.~Yu, M.~L. Hu, C.~X. Zhang, C.~Y. He, L.~Z. Sun, J.~Zhong, J of Physical
  Chemistry C 115 (2011) 10836.

\bibitem{peng12}
Q.~Peng, S.~De, Physica E 44 (2012) 1662.

\bibitem{jena11}
M.~Kan, J.~Zhou, Q.~Wang, Q.~Sun, P.~Jena, Phys. Rev. B 84 (2011) 205412.

\bibitem{grossman12}
M.~Bernardi, M.~Palummo, J.~Grossman, Phys. Rev. Lett 108 (2012) 226805.

\bibitem{cnrrao13}
N.~Kumar, K.~Moses, K.~Pramoda, S.~N. Shirodkar, A.~K. Mishra, U.~V. Waghmare,
  A.~Sundaresana, C.~N.~R. Rao, J. Mater. Chem. A 81 (2013) 109.

\bibitem{ashcroft76}
N.~W. Ashcroft, N.~Mermin, Solid State Physics, Holt, Reinhart and Winston, New
  York, 1976.

\bibitem{madsen06}
G.~Madsen, D.~Singh, Computer Physics Communications 175 (2006) 67.

\bibitem{giannozzi09}
P.~Giannozzi, et~al., J. Phys. Condens. Matter 21 (2009) 395502.

\bibitem{vanderbilt90}
D.~Vanderbilt, Phys. Rev. B 41 (1990) 7892.

\bibitem{pbe96}
J.~P. Perdew, K.~Burke, M.~Ernzerhof, Phys. Rev. Lett. 77 (1996) 3865.

\bibitem{mp76}
H.~J. Monkhorst, J.~D. Pack, Phys. Rev. B 13 (1976) 5188.

\bibitem{soler95}
D.~Sanchez-Portal, E.~Artacho, J.~M. Soler, Sol. St. Commun 95 (1995) 685.

\bibitem{xcrysden03}
A.~Kokalj, Comp. Mater. Sci. 28 (2003) 155.

\bibitem{allen88}
P.~Allen, W.~Pickett, H.~Krakauer, Phys. Rev. B 37 (1988) 7482.

\bibitem{allen92}
W.~Schulz, P.~Allen, N.~Trivedi, Phys. Rev. B 45 (1992) 10886.

\bibitem{sm11}
T.~P. Kaloni, S.~Mukherjee, Modern Physics Letters B 25 (2011) 1855.

\bibitem{charge}
The excess charge, calculated on the inequivalent atoms in C$_x$(BN)$_{1-x}$,
  the spin polarization on each atom in carbon nanoribbon and the free energy
  showing the binodal to spinodal transition for armchair and zigzag interfaces
  are given in the Supplementary informations.

\bibitem{zunger87}
J.~E. Bernard, A.~Zunger, Phys. Rev. B 36 (1987) 319.

\bibitem{lambrecht93}
W.~R.~L. Lambrecht, B.~Segall, Phys. Rev. B 47 (1993) 9289.

\bibitem{neugebauer95}
J.~Neugebauer, C.~G. {Van de Walle}, Phys. Rev. B 51 (1995) 10568.

\bibitem{liou05}
B.~T. Liou, S.~H. Yen, Y.~K. Kuo, Applied Physics A 81 (2005) 651.

\bibitem{hse}
J.~Heyd, G.~E. Scuseria, M.~Ernzerhof, J Chem Phys 118 (2003) 8207.

\bibitem{gw}
L.~Hedin, Phys Rev 139 (1965) A796.

\end{thebibliography}
\end{document}